\documentclass[11pt]{article}
\usepackage[pctex32]{graphics}
\usepackage{amsmath,amsthm}
\usepackage{amssymb,latexsym}
\usepackage[mathscr]{eucal}

\usepackage{setspace}

\newcommand{\bee}{\begin{equation}}
\newcommand{\ene}{\end{equation}}

\begin{document}

\begin {center} {\bf Critical Exponents and Universality in Fully Developed Turbulence}

\vspace{.3in}

Bhimsen K. Shivamoggi\\
University of Central Florida\\
Orlando, FL 32816-1364
\end{center}

\vspace{.5in}

{\bf Abstract}. -  Multi-fractal model for hydrodynamic fully-developed turbulence (FDT) has been used to provide a detailed structure for the critical exponent $\sigma$ describing the scaling form of energy (or enstrophy) dissipation rate $\epsilon$ (or $\tau$) that appears to exhibit an interesting {\it universality} covering radically different hydrodynamic FDT systems. This result also appears to provide a consistent framework for classification of dissipation field into {\it critical}, {\it subcritical} and {\it supercritical} cases. Some FDT problems that exemplify these cases are discussed.

\medskip


\medskip

{\bf Introduction}. - Small-scale structure in three-dimensional (3D) incompressible fully-developed turbulence (FDT), following Kolmogorov's [1] epoch-making work, is believed to possess, in the large Reynolds number ($R\Rightarrow\infty$) limit, a certain universality in its scaling properties. This universal scaling behavior depends only on symmetries and conservation laws of the system and is unaffected by the large-scale flow structure.

This is reminiscent of the scaling behavior near the critical point where many diverse systems show a striking similarity in their behavior [2,3].  Critical phenomena had a theoretical breakthrough in the renormalization group (RG) [4,5] which was the culmination of the ideas of scaling and universality. The formal application of RG procedure was attempted for the FDT problem [6-9]. The goal is to determine the critical exponents (like those associated with correlation length in critical lattice spin systems)  that are intrinsic features of the FDT dynamics and not artifacts of the large-scale stirring mechanisms and hence unify radically different FDT systems near their critical points.

Spatial intermittency is a common feature of FDT and implies that turbulence activity at small scales is not distributed uniformly throughout space. This leads to a violation of an assumption in the Kolmogorov [1] theory that the statistical quantities show no dependence in the inertial range $L\gg\ell\gg\eta$ on the large scale $L$ (where the particular external stirring mechanisms generating FDT become influential) and the Kolmogorov microscale $\eta =(\nu^3 /\epsilon)^{1/4}$ (where the viscous effects become important). Thus, if one views the Kolmogorov [1] theory  as a mean field theory [10], the spatial intermittency aspects will be expected to define at least one additional universal scaling exponent [11]. Spatial intermittency effects can be very conveniently imagined to be related to the fractal aspects of the geometry of FDT [12]. The mean energy dissipation field $\epsilon$ may then be assumed, in a first approximation, to be a homogeneous fractal [13], and more generally,  a multi-fractal [14-17]. The latter idea has received experimental support [18].

In the $R\Rightarrow\infty $ limit (which corresponds to the critical point for FDT) infinitely many length scales become important. So, by analogy with critical phenomena, one may expect many macroscopic details  to become irrelevant and the critical behaviors of radically different FDT systems to exhibit some universality. Further, this universality may be expected to be strongly rooted in the self-similarity of the system at the critical point just as that for a finite critical ferromagnet system  via the multi-fractal structure of the $\epsilon$-field [19].

One of the fundamental dynamical assumptions underlying FDT theory is that  $\epsilon$ remains finite in the $R\Rightarrow\infty $ limit (see  [20] for experimental support for this assumption [21] -- the infinite $R$ limit here is taken to imply $\nu \Rightarrow 0$, see (5) below). This implies that if one identifies $\epsilon$ as the order parameter, then the critical exponent $\sigma$ defined by
\bee
\epsilon\sim (\bar R)^\sigma , \quad \bar R\Rightarrow\infty 
\ene
is
\bee
\sigma =0. 
\ene

This aspect has also been validated by the multi-fractal model for the  $\epsilon$-field [23]. The purpose of this letter is to report that the multi-fractal model indeed provides a detailed structure for the critical exponent $\sigma$ that appears to exhibit an interesting {\it universality} covering radically different FDT systems.

{\bf The Multi-fractal Model for Classical 3D FDT}. - In the multi-fractal model for the classical 3D FDT system, one stipulates that the fine-scale regime of FDT possesses a range of scaling exponents $\alpha \in I\equiv [\alpha_{\min} ,\alpha_{\max}]$. Each $\alpha\in I$ has the support set $S(\alpha )\subset\mathbb R^3$ of fractal dimension $f(\alpha )$ [24] such that, as $\ell\Rightarrow 0$, the velocity increment has the scaling behavior 
\bee
\delta V(\ell )\sim\ell^\alpha .
\ene
The sets $S (\alpha )$ are nested so that $S(\alpha')\subset S(\alpha )$, for $\alpha'<\alpha$.

On extrapolating this multi-fractal scaling in the inertial range down to the Kolmogorov microscale $\eta$ [25], the latter is found to exhibit the scaling behavior [26] 
\bee
\left(\frac{\eta}{L}\right) \sim (\bar R_1)^{-\frac{1}{1+\alpha}}
\ene
where $\bar R_1$ is a mean Reynolds number,
\bee
\bar R_1 \sim\frac{(\bar \epsilon L^4)^{1/3}}{\nu} ,
\ene
$\bar \epsilon$ being the mean energy dissipation rate.

The moments of the velocity-gradient distribution are then given by 
\bee
A_p^{(1)} \equiv\left< \left|\frac{\partial v}{\partial x} \right|^p\right> \equiv\int (\bar R_1)^{-\frac{1}{1+\alpha} [p\alpha -p+3-f(\alpha )]} d\mu (\alpha )
\ene
In order to determine the scaling behavior of $A_p^{(1)}$, first note that the dominant contribution to the integral in (6), in the limit $\left< R_1\right> \Rightarrow \infty$, corresponds via method of steepest descent to
\bee
\frac{d}{d\alpha} \left[ \frac{p\alpha -p+3-f(\alpha )}{1+\alpha} \right] \Bigg|_{\alpha =\alpha_*} =0.
\ene

Next, the sums of the moments of the total energy dissipation $E(\ell )\sim\epsilon (\ell )\ell^3$ occuring in $N(\ell )$ boxes of size $\ell$ covering the support of the measure $\epsilon$ exhibit the following asymptotic scaling behavior [27]
\bee
\sum_{i=1}^{N(\ell )} [E_i(\ell )]^q \sim \int \ell^{(3\alpha +2)q-f(\alpha )} d\mu (\alpha )\;\sim \;\ell^{(q-1)D_q} , 
 \quad \ell \;\;\text{small},
\ene
$D_q$ being the generalized fractal dimension (GFD) of the $\epsilon$-field [28]. The dominant contribution to the integral in (8), in the limit $\ell \Rightarrow 0$, corresponds again via method of steepest descent to
\begin{subequations}
\bee
(3\alpha_*+2)q-f(\alpha_*)=(q-1)D_q
\ene
where,
\bee
\frac{df(\alpha_*)}{d\alpha} =3q.
\ene
\end{subequations}

The coincidence of the values of $\alpha_*$ given by (7) and (9b) for which the integrands in (6) and (8) become extremum is insured by assuming the Kolmogorov refined similarity hypothesis [18] in the Kolmogorov microscale regime.

Combining (8) and (9), we obtain  -
\bee
A_p^{(1)} \sim (\bar R_1)^{-\frac{D_Q (p-3)-5p+9}{D_Q+1}} , \; \bar R_1\Rightarrow\infty
\ene
where $Q$ is the root of
\bee
Q=\frac{D_Q+2p-3}{D_Q+1} .
\ene

Using (11), (10) may be rewritten as
\begin{subequations}
\bee
A_p^{(1)} \sim (\bar R_1)^{\gamma_p^{(1)}} , \;\; \bar R_1\Rightarrow\infty
\ene
where,
\bee
\gamma_p^{(1)}\equiv -(p-3Q).
\ene
\end{subequations}
Thus, the mean energy dissipation $\nu A_2^{(1)}$ has the following scaling behavior -
\begin{subequations}
\bee
\nu A_2^{(1)} \sim (\bar R_1)^{\sigma_1} , \;\; \bar R_1 \Rightarrow\infty
\ene
where,
\bee
\sigma_1\equiv 3(Q-1).
\ene
\end{subequations}

On the other hand, corresponding to $p=2$, (11) yields
\bee
Q=1
\ene
in agreement with [23]. Using (14), (13b) leads to
\bee
\sigma_1 =0
\ene
and (13a) then leads to
\bee
\nu A_2^{(1)} \sim (\bar R_1)^0 \sim \text{constant}, \;\; \bar R_1\Rightarrow\infty
\ene
validating the inviscid dissipation of energy in classical 3D FDT! Further, (14) implies that the $\epsilon$-field in classical 3D FDT has the GFD $D_Q$ {\it equal} to the {\it information entropy} dimension $D_1$.

{\bf Applications to Other FDT Systems}. - Let us consider next the classical two-dimensional (2D) incompressible FDT system [29]. Following the above procedure, the scaling behavior of
the mean enstrophy dissipation $\nu A_2^{(2)}$ (which is the relevant physical quantity in the enstrophy cascade of 2D FDT) can be calculated from the results in [30] -
\begin{subequations}
\bee
\nu A_2^{(2)} \sim\left< \left| \frac{\partial^2v}{\partial x^2} \right|^2\right> \sim ( \bar  R_2)^{\sigma_2} , \;\;\bar R_2 \Rightarrow\infty
\ene
where,
\bee
\sigma_2\equiv 3(Q-1),
\ene
\end{subequations}
and
\bee
\bar R_2\sim\frac{(\bar \tau L^6 )^{1/3}}{\nu}
\ene
$\bar \tau$ being the mean enstrophy dissipation rate, and $Q$ is the root of -
\bee
Q=\frac{D_Q+3p-2}{D_Q+4} .
\ene
$\sigma_2$ is identical to the exponent $\sigma_1$ (13b) for the mean energy dissipation in 3D FDT! Corresponding to $p=2$, (19) yields
\bee
Q=1.
\ene
Using (20), (17b) leads to
\bee
\sigma_2\equiv 0
\ene
and (17a) then leads to
\bee
\nu A_2^{(2)} \sim (\bar R_2)^0 \sim \text{constant}, \;\; \bar R_2 \Rightarrow\infty
\ene
establishing the inviscid dissipation of enstrophy in the enstrophy cascade of classical 2D FDT. 

Let us consider next 3D compressible (barotropic fluid) FDT [31]. The scaling behavior of the mean kinetic energy dissipation $\mu A_2^{(1)}$ can be calculated from the results in [31] -
\bee
\mu A_2^{(1)} \sim (\bar R_1 )^{\sigma_3} , \quad \bar R_1 \Rightarrow\infty
\ene
where,
\vspace*{-.1in}
\bee
\sigma_3\equiv\left(\frac{3\gamma -1}{\gamma +1}\right) (Q-1)
\ene
and $Q$ is the root of
\bee
Q=\frac{D_Q+\left(\frac{2\gamma}{\gamma +1}\right)p-3}{D_q+\left(\frac{\gamma -3}{\gamma +1}\right)}
\ene
$\gamma$ being the ratio of specific heats of the fluid. (24) exhibits the universality but with a different amplitude. The latter reflects the residual effect of variations in the cascade physics involving the acoustic wave dynamics in the compressible case [32]. Corresponding to $p=2$, (25) yields
\bee
Q=1.
\ene
Using (26), (24) yields
\vspace*{-.1in}
\bee
\sigma_3=0
\ene
and (23) leads to
\bee
\mu A_2^{(1)}\sim (\bar R_1)^0 \sim \text{ constant}, \;\; \bar R_1 \Rightarrow\infty
\ene
establishing the inviscid dissipation of kinetic energy in 3D compressible FDT.

These results imply that energy cascades of 3D and the enstrophy cascade of
2D FDT are indeed examples of the {\it critical} dissipation field which -
\begin{itemize}
\item has the GFD {\it equal} to the {\it information entropy} dimension $D_1$,
\vspace*{-.1in}
\item has a {\it finite} mean value in the inviscid limit.
\end{itemize}
This suggests further that we may define a {\it subcritical/supercritical} dissipation field as one which -
\begin{itemize}
\item has the GFD {\it less/greater} than the {\it information entropy} dimension $D_1$,
\vspace*{-.1in}
\item has an {\it infinite/vanishing} mean value in the inviscid limit;
\end{itemize}

As an example of a {\it subcritical} dissipation field, consider the 2D quasi-geostrophic FDT system in which the baroclinic effects are produced by the deformed free-surface of the ocean [30]. The scaling behavior of the  mean enstrophy dissipation $\nu A_2^{(2)}$ can again be calculated from the results in [30] -
\begin{subequations}
\bee
\nu A_2^{(2)} \sim (\bar R_2 )^{\sigma_3} , \;\; \bar R_2\Rightarrow\infty
\ene
where,
\bee
\sigma_3\equiv 3(Q-1),
\ene
\end{subequations}
and $Q$ is the root of
\bee
Q=\frac{3D_Q+11p-6}{3(D_Q+4)} .
\ene
$\sigma_3$ is again identical to the corresponding exponents for the classical 3D and 2D FDT! Corresponding to $p=2$, (30) yields,
\bee
Q=1+\frac{4}{3D_Q+12} >1.
\ene
Using (31), (29b) leads to
\bee
\sigma_3>0
\ene
and (29a) then leads to
\bee
\nu A_2^{(2)} \Rightarrow\infty , \;\; \bar R_2 \Rightarrow\infty .
\ene
So, the enstrophy dissipation in the enstrophy cascade for the 2D quasi-geostrophic FDT system diverges [36] in the inviscid limit in accordance with the first of the above stipulations for the {\it subcritical} dissipation field. Further, (31) implies that the enstrophy dissipation field has the GFD $D_Q$ {\it less} than the {\it information entropy} dimension $D_1$ ($D_Q$ can be shown [37] to be monotonically decreasing with $Q$), in accordance with the second of the above stipulations for the {\it subcritical} dissipation field.

{\bf The Probability Density Functions of the Flow-Varaible Gradients}. - The universality of the spatial intermittency aspects in FDT systems, on the other hand, would imply that the {\it criticality}, {\it subcriticality} or {\it supercriticality} of the dissipation field should be traceable to properties like the probability density function (PDF) in the {\it zero} intermittency limit of the FDT system in question.

The PDF of the velocity gradient in the {\it zero} intermittency limit for the classical 3D FDT is given by (Frisch and She [38])
\bee
P(s)\sim\left(\frac{\nu}{|s|}\right)^{1/3} e^{- {\displaystyle \left[  \frac{\nu^{2/3}|s|^{4/3}}{2\left< v_0^2\right>}\right]}}
\ene
where,
$$
s\sim\frac{\partial v}{\partial x}
$$
and $v_0$ is the velocity increment characterizing large scales.

The PDF of the vorticity gradient in the {\it zero} intermittency limit for the classical 2D FDT is given by (Shivamoggi [30])
\bee
P(r)\sim\left(\frac{\nu}{|r|}\right)^{1/3} e^{- {\displaystyle \left[\frac{\nu^{2/3}|r|^{4/3}}{2\left< v_0^2\right>}\right]}}
\ene
where,
$$
r\sim\frac{\partial^2 v}{\partial x^2} \, .
$$

The PDF of the velocity gradient in the {\it zero} intermittency limit for 4D compressible FDT is given by (Shivamoggi [31])
\bee
P(s)\sim\left(\frac{\nu_0}{|s|}\right)^{1/3} e^{- {\displaystyle \left[\frac{\nu_0^{2/3}|s|^{4/3}}{2\left< v_0^2\right>}\right]}}
\ene
where $\nu_0$ is a reference kinematic viscosity.

The identity of (34)-(36) appears to confirm that energy cascades of 3D and the enstrophy cascade of 2D FDT are examples of the {\it critical} dissipation field.

On the other hand, the PDF of the vorticity gradient in the {\it zero} intermittency limit for the 2D geostrophic FDT is given by (Shivamoggi [30])
\bee
P(r)\sim\left(\frac{\nu}{|r|}\right)^{5/11} e^{- {\displaystyle \left[\frac{\nu^{10/11}|r|^{12/11}}{2\left< v_0^2\right>}\right]}}
\ene
which is more non-Gaussian than (35) and is consistent with the fact that 2D geostrophic FDT is an example of a {\it subcritical} dissipation field.

{\bf Discussion}. - According to the multi-fractal model, the order parameter $\epsilon$ (or $\tau$) in a variety of hydrodynamic FDT cases therefore has a scaling form
\begin{subequations}
\bee
\epsilon \sim (\bar R)^\sigma , \;\; \bar R\Rightarrow\infty
\ene
with the apparently {\it universal} critical exponent  $\sigma$ given by
\bee
\sigma =a(Q-1)
\ene
\end{subequations}
$D_Q$ being the GFD of the dissipation field. $Q \lesseqgtr 1$ corresponds to {\it supercritical}, {\it critical} (of which classical 3D and enstrophy cascade of 2D FDT are examples), and {\it subcritical} (of which 2D geostrophic FDT is an example) dissipation-field cases. On first thought, the {\it universality} of critical exponent $\sigma$ would appear to be rather far-fetched because the various cases of FDT are far too diverse to be characterized within a universality class. But, as speculated in [19], by focusing on the critical point $(R\Rightarrow\infty )$ behavior, it is apparently possible to do just that. 

I am thankful to Professors John Cannon, Greg Eyink, Grisha  Falkovich,  Siegfried Grossmann,  Mike Johnson, David Kaup, Evgeny Novikov, Katepalli Sreenivasan, Henri Tasso, Angelo Vulpiani and Victor Yakhot for their helpful comments and advice.



\begin{thebibliography}{99}
\bibitem{} The multi-fractal  model was shown [17] to be a version of the scale-similarity of the so-called breakdown coefficients.
\vspace*{-.1in}
\bibitem{} This assumption has been  rigorously shown from Navier-Stokes equations to hold as an upper bound on $\epsilon$ [22].
\vspace*{-.1in}
\bibitem{} The function $f(\alpha )$ is postulated to be universal, i.e, independent of the mechanism of generation of FDT. Further, $f(\alpha )$ can be determined directly in experiments [18].
\vspace*{-.1in}
\bibitem{} This is physically plausible because, thanks to the vorticity enhancement produced by the deformed free surface, quasi-geostrophic flows, in the baroclinic limit, are more dissipative than the classical 2D flows [30].
\vspace*{-.1in}
\end{thebibliography}
\end{document}